\documentclass[manuscript]{aastex}
\usepackage{graphicx}
\usepackage{epstopdf}
\usepackage{color}
\definecolor{Blue}{rgb}{0.3,0.3,0.9}

\shorttitle{Atmospheric Reflection in Hercules X-1}
\shortauthors{M. H. Abdallah et al.}

\begin{document}

\title{Spectral Signature of Atmospheric Reflection \\
    in Hercules X-1/HZ Hercules During Low and Short-High States}

\author{M. H. Abdallah\altaffilmark{1} and D. A. Leahy}
\affil{Dept. of Physics and Astronomy, University of Calgary,
    Calgary, Alberta, Canada T2N 1N4}
\email{mhabdall@ucalgary.ca}

\altaffiltext{1}{Astronomy Department, National Research Institute of Astronomy and Geophysics (NRIAG), 11421 Helwan, Cairo, Egypt}

\begin{abstract}

We analyse RXTE/PCA X-ray spectra of the binary X-ray pulsar Her X-1/HZ Her 
during short high state and one binary orbit in the preceding low state, just before short high turn-on.
The spectrum is well described by two continuum components (absorbed and unabsorbed).
The resulting spectral parameters are modulated with orbital phase.  
During low state a significant component of the flux, and its spectrum, is consistent with 
X-ray reflection off the face of the companion star HZ Her. This component has a significantly
harder X-ray spectrum than the rest of the flux from the Her X-1 system. 
A second component in low state is consistent with emission from the accretion disk corona.
During short high a third strong component is present with a softer spectrum, 
which is associated with the neutron star and accretion disk.
Due to this direct emission from the neutron star and accretion disk, the reflected emission 
is less clear, however parameters and fluxes modulations during short high state indicate its presence. 
In low state, the hard X-ray flux ($h \nu$ $>$ $10 keV$) peaks at  
orbital phase $\phi_{orb}\simeq$ 0.55, which is expected 
from a simple model of atmospheric reflection from the companion star. 
The offset indicates an asymmetry in the X-ray illumination of the companion, which 
could be due to shadowing of the the inner face of HZ Her by the accretion disk and/or stream.

\end{abstract}


\section{Introduction}

Hercules X-1 (Tananbaum et al. 1972) is one of the brightest, and most studied accreting 
X-ray binaries. 
It consists of a neutron star, Hercules X-1 (Her X-1) with rotation period of 1.23-s, 
in an orbit of 1.7-day (Deeter, Boynton $\&$ Pravdo 1981) around a 
A/F stellar companion HZ Herculis (HZ Her) (Doxsey et al. 1973). 
The pulsar and the companion masses are 1.5 $M_{\bigodot}$ and 2.2 $M_{\bigodot}$ 
respectively (Leahy and Abdallah, 2014, Reynolds et al 1997) making it the only known 
Intermediate-Mass (IMXB) system, resembling the High-Mass (HMXB) systems by 
its pulsation and the domination of optical emission by the companion, 
while matter transfer via Roche lobe overflow creates the accretion disk as in 
Low-Mass (LMXB) systems. 
In addition to the rotational and orbital periods, the system 
exhibits a 35-day super-orbital cycle, 
of high and low X-ray flux states consisting of: Main-High (MH) state covering 
phases 0-0.31, Short High (SH) state covering phases 0.57-0.79 with flux peaking 
at $\sim$ $30\%$ of the MH maximum, and two Low States (LS), with flux that remains at $\sim$ $1\%$ of 
the maximum, separating the high states (Giacconi et al 1973; Scott and Leahy 1999; Igna and Leahy 2011).

\vspace{10 mm}

Due to the high orbital inclination of the binary ($i$ $ \gtrsim $ $80^{\circ}$), 
this super-orbital cycle is interrupted once per orbital period by the eclipse of the 
neutron star by its companion. 
The eclipse range is 0.93-0.07 in orbital phase (Scott and Leahy 1999). 
Absorption events, termed X-ray dips, are also frequently observed because of the 
high inclination of the system. 
These dips occur throughout the orbital cycle and the 35-day cycle high states. 
The dip classification in orbital phase is based on the following phase intervals: 
0.07-0.45 for post-eclipse dips; 0.45-0.65 for anomalous dips; and 0.65-0.93 for pre-eclipse
 dips (Giacconi et al. 1973; Gerend and Boynton 1976; Scott and Leahy 1999; Igna and Leahy 2012). 
 The 35-day periodicity in the system is the signature of periodic obscuration of X-ray 
 emission from the central neutron star by the precession of a tilted and twisted accretion 
 disk viewed nearly edge on (Petterson 1975; Wijers and Pringle 1999; Scott et al. 2000; Leahy 2002). 
The misalignment between the magnetic and rotational axes of the neutron star  gives the  
distinctive pulse profile (Leahy 2004) which evolves over the 35-day 
cycle (Deeter et al. 1998; Scott, Leahy, and Wilson 2000). 
There are times when the system enters into a long period of low X-ray 
flux state, referred to as Anomalous Low States (ALS) (Parmar et al. 1985; 
Vrtilek et al. 1994; Oosterbroek et al. 2000). 
The cause of these ALS is not clear, but they probably result from changes 
in the state of the accretion disk (Leahy and Dupuis 2010).

\vspace{10 mm}

The system is located at high galactic latitude with low interstellar hydrogen column density
 making Ultraviolet (UV) and Extreme Ultraviolet (EUV) observations feasible. 
 The accretion disk casts a shadow that migrates across the surface of the companion star on the 
 beat frequency between the orbital and 35-day cycles. 
This beat frequency cycle is consequently observed at UV and optical energies, 
which are dominated by X-ray emission reprocessed in the stellar atmosphere 
(Gerend and Boynton 1976, Leahy and Marshall 1999). 
In close binary systems some of X-ray radiation reaching the surface 
of the companion star gets reflected (\textsection 2). 
This phenomena is easily seen in X-ray light curve when the direct emission 
from the X-ray pulsars is hidden from the observer. 
In the present paper, our main focus is on the spectral modulation with orbital 
phase of the system Her X-1/HZ Her due to reflected X-ray emission from the atmosphere of HZ Her. 
This X-ray modulation has been clearly observed during LS and for the first time, 
also detected in SH-state.

\section{Atmospheric Reflection}

In X-ray binaries, the irradiated cold atmosphere of the companion star 
will scatter some of the incident hard radiation ($h\nu \gtrsim 10$ keV), 
while the remainder is absorbed and heats the stellar photosphere, 
emerging as Optical/UV radiation. 
The scattered radiation is termed X-ray reflection. 
Lower energy photons ($h \nu \lesssim 8$  keV) are mainly absorbed through the 
photoionization of the K-shell electrons of heavy elements and transformed into optical emission. 
Because of the rapid decrease of of photoionization
cross-section with with increasing frequency $ \sigma_{ph.} (\nu) \varpropto \nu^{-3}$, the 
photoabsorption becomes ineffective for high energy photons and the Thomson scattering cross-section 
exceeds the photoionization cross-section for high energy photons ($h \nu \gtrsim 8$  keV). Thus
the reflected spectrum is harder than the that of the primary emission. 
X-rays interacting with the companion's atmosphere also induce extra stellar wind 
(Basko and Sunyaev 1973; Arons 1973) creating a high temperature layer of highly ionized 
heavy elements in which the photo-absorption optical depth becomes small for 
photons with $h\nu \gtrsim 2$ keV. 
The Thomson scattering optical depth in this zone can be significant, $\tau_{\circ} \sim 0.05$
 for the system HZ Her/Her X-1 (Basko and Sunyaev 1973), causing reflection of soft X-rays 
($h\nu \lesssim 10$ keV).
The photoabsorption of hard X-ray photons by heavy elements Fe, S, and Ar are transformed 
into $K_{\alpha}$ emission. Iron $K_{\alpha}$ line is the most likely to be observed because the K-fluorescence 
yield $\omega_{k}$ is proportional to the fourth 
power of nuclear charge $Z^{4}$. Thus the reflected spectrum characterized by 
a strong iron $K_{\alpha}$ emission line (see Figure 2 of Basko et al. 1974).
When the radius of the normal star is comparable to $cp$ , where $c$ is the speed of light and $p$ is the pulse period, the 
pulses will be blurred as the reflected emission from different parts of the photosphere will reach
the observer at different phases of the pulse. 
 The solid angle subtended by HZ Her at the X-ray source is 
$\sim$ 0.6 sr thus the reflected X-ray flux should be overwhelmed by the direct emission 
from the neutron star during the high states. 
However, during the low states when the direct emission is obscured by the accretion disk, 
the system should be observed at 5-10\% of the direct flux (Basko et al. 1974). 
The reflected X-ray spectra from a planar atmosphere has been studied by Basko et al. 1974. 
They found that $30-40\%$ of the incident X-ray in the 2-30 keV energy range irradiating the face of HZ Her is reflected. 
Felsteiner and Opher 1976 conducted a similar study and found significant but smaller albedo than Basko et al. 1974.

\vspace{10 mm}

The amount of the reflected radiation from HZ Her decreases with increasing the metal ($Z \geq 3$) abundances. 
Figure 2 of Bai (1980) shows the reflected continuum at orbital phase $\phi_{orb} = 0.5$ 
for the cases of metal abundance of one, two, and four times solar values. 
The reflected spectrum is characterized by a strong iron $K_{\alpha}$ emission line. 
Iron line transfer in the stellar atmosphere of a normal star illuminated by an external 
X-ray source has been studied by Hatchett and Weaver (1977). 
They found that the magnitude of reflected line flux depends on the incident X-ray spectrum 
and on the solid angle subtended by the companion at the X-ray source, 
but is insensitive to the abundance of heavy elements. 

To summarize, the predicted characteristics of the reflected flux from the HZ Her/Her X-1 system are as follows:
 (i) it should be easily detected during the low-states as its amplitude is about 10\% of the 
 incident flux; 
 (ii) the intensity should vary with orbital phase because of the changing visibility of the irradiated face of HZ Her; 
 (iii) X-ray pulsations should be suppressed; 
 (iv) the maximum and minimum of the reflected X-ray light curve should coincide with those of optical/infra-red light curves; 
 (v) the reflected spectrum should be harder than that of the primary flux; 
 (vi) there should be a strong iron $K_{\alpha}$ emission line.

\vspace{10 mm}

The reflection effect in HZ Her/Her X-1 system has been detected in the EUV 
energy band using Extreme Ultraviolet Explorer (EUVE) 
Deep Survey (DS) instrument with lexan/B filter (Leahy \& Marshall 1999 ; Leahy et al. 2000). 
Leahy \& Marshall (1999) observed Her X-1 at 35 day phase 0.76-0.88, which covers the end of the SH-state and the LS. 
The count rate during this observation peaked at $\sim$ 0.02 counts/s and was strongly modulated at the binary period. 
They concluded that most of the observed EUV emission was reflected emission from the companion star HZ Her. 
EUV observations at 35 day phase 0.5-0.62 covering two complete orbital cycles including 
the turn-on to the SH-state of Her X-1 also show orbital modulation both before and after the turn-on (Leahy et al. 2000). 
They found that The peak count rate of the LS part before the turn-on of the SH-state is approximately at twice
that of the LS after SH-state.
They attributed a change in orbital modulation to the difference in the shadowing of HZ Her by 
the accretion disk, and constructed a model for the disk shadowing (see their figure 5). 
They calculated the reflected intensity from HZ Her for face-on geometry (orbital phase 0.5) as 
a function of 35 day phase/disk phase (see their Figure 6) and found a modulation of reflected intensity 
by a factor of $\sim$ 2.7 over disk phase.

\vspace{10 mm}

X-ray orbital modulation of the HZ Her/Her X-1 binary system was clearly detected during the 
ALS (Still et al. 2001), for data covering over 2.7 orbital cycles. 
The X-ray light curve during this observation resembles that of optical and UV light curves, 
attaining a maximum when the irradiated face of HZ Her is most visible. 
Still et al. (2001) interpreted the modulation as caused by a change with orbital phase 
in X-rays reflected off the atmosphere of the companion star. 
In observation of Her X-1 during ALS by ASTRON mission (Sheffer et al. 1992), 
the observed flux was at $\sim$ 5\% of the maximum. 
This data was take at orbital phase $\phi_{orb} = 0.525$. 
They found two emission components (hard \& soft components) in their spectrum 
corresponding to reflection from the photosphere of HZ Her, and scattering by the corona 
of the star/disk respectively. 
However they didn't have enough orbital phase coverage to see the modulation expected for 
reflected emission. 
A similar spectrum during normal LS was seen by Mihara et al. (1991) but due to the 
lack of orbital modulation they interpreted the hard component as forward scattered of the main-on emission. 
However their data only covered 0.41-0.50 \& 0.56-0.60 in orbital phase. 
Modulation of the Fe $K_{\alpha}$ line with binary phase was observed by XMM-Newton 
during LS (Zane et al. 2004).
However because of the fast rise of line flux with orbital phase, it was difficult to 
associate this modulation with the companion star. 
Their data covered a wide range in 35-day phase during low states from five different 35-day cycles and 12-orbital periods. 
This might have confused the orbital modulation of the line flux because of changes in the 
disk emission and changes in the shadowing of the companion by the disk with 35-day phase 
similar to that found for EUV emission (Leahy \& Marshall 1999 ; Leahy et al. 2000).

\section{Observations and Results}

The results presented here are based on data taken by the proportional counter array (PCA) 
(Jahoda et al. 1996; Jahoda et al. 2006) on RXTE. 
The PCA instrument consists of an array of five nearly identical Xenon Proportional Counting 
Units (PCUs) operating independently in the 2-60 keV energy range with a moderate energy resolution of $18\%$ at 6 keV. 
Each detector is made of multi-wire, multi-layer, proportional counter with a collecting area of 1600 $cm^{2}$. 
The number of operating PCUs varied between 2 and 5 during the observations as a 
standard procedures to extend the lifetime of the instrument. 
The observations were made in the period from 18 to 26 of December 2001 (MJD 52261 - 52269) 
and having a common proposal ID 60017. 
The observation contains $\sim$ 3.6 binary orbits in SH-State and one orbit 
in the preceding LS, just before the turn-on.
Dips significantly affect the spectral parameters, and would interfere with the 
search for spectral change due to atmospheric reflection by HZ Her during SH-state.
There is no reflected emission by the companion during eclipse. 
Thus the data used for spectral analysis here omit eclipse and dips times.
The resulting total eclipse-and-dips-free exposure is 117 ksec. 
 
\vspace{10 mm}

Figure (1) shows the band-1 (2-4 keV), band-3 (9-20 keV) light curves, and the band-1/band-3 softness ratio. 
For these light curves we averaged the count rates over time intervals between 112-s and 1680-s with 
a mean value of $\sim$ 1 ks .
The count rates are scaled by the number of active PCUs to yield count rate per PCU. 
A smooth orbital modulation in count rate is seen during LS as 
expected from the changing of the reflected emission off the companion atmosphere.
The count rate in Band-3 steadily increases from 1.29 c/s to 4.72 c/s with a 
maximum attained at orbital phase $\phi_{orb}$ $=$ 0.54 and then decreases slowly during the second half of orbital period. 
Similar modulations in band-3 are also seen in SH-State. 
These are most clearly visible in band-3 between MJD 52264.8 \& 52265.8. 
To search for similar modulations of various spectral parameters, the standard2 data were used for spectral analysis. 
Data were first filtered to reject the following periods: 
i) pointings closer than $10^{\circ}$ to the Earth's limb, to avoid X-ray contamination from the Earth; 
ii) data from a 30 minute interval beginning with the satellite entering the SAA; 
iii) times in which the electron ratio in each of the accumulated PCUs larger than 0.1, to avoid electron contamination; 
iv) periods of unstable pointing (OFFSET $\geq$ $0^{\circ}.02$ from the source position). 
Energy spectra in 129 channels ,covering energy range 2-60 keV were extracted with exposure times between 112-s and 1680-s with a mean value of 998-s. 
This resulted in total of 117 spectra: 41 spectra during LS, and 76 during SH-state. 
Data from all layers and columns of the available PCUs were added together and response matrices were generated accordingly. 
The background estimation was done using the RXTE/PCA faint model released by the PCA instrument team. 

\vspace{10 mm}

We used XSPEC (v.12.7.1) for fitting the spectra in the energy range 2.5-30 keV. 
The first three energy channel were ignored because of uncertain background modelling, 
while channels 59-129 ($>$ 30 keV) were ignored because of poor counting statistics. 
In the 2.5-30 keV energy range the spectrum of Her X-1 is best described by a two 
component power-law continuum (absorbed and unabsorbed) model with high-energy exponential 
cut-off throughout the 35-day cycle (Mihara et al. 1991, Leahy et al. 1991, Leahy 2001). 
These two spectral components represent various physical emission components in the binary (Figure 2).
We used a power-law model with high energy exponential cutoff (powerlaw/highecut in XSPEC), a Gaussian for Fe K-emission 
line, and partial covering absorption (pcfabs in XSPEC). 
To reduce residuals at the cutoff energy due to the discontinuity at this energy we 
applied a smoothing Gaussian-shaped absorption function (gabs in XSPEC) at the cutoff energy 
with width and depth fixed at 1.5 keV and 0.35 respectively (see Coburn et al. 2002). 
The model used take the following form:

\vspace{10 mm}
\begin{equation}
I(E) = pcfabs \times gabs \times I_{cutoffpl}(E) + I_{Fe}(E) 
\end{equation}
where,

\begin{eqnarray}
pcfabs = F \times exp[-n_{H} \sigma(E)] + (1 - F) \nonumber
\end{eqnarray}

\begin{eqnarray}
gabs = exp[-(\tau_{g}/\sqrt{2\pi} \sigma_{g}) exp(-0.5((E - E_{cut})/\sigma_{g})^{2})] \nonumber
\end{eqnarray}


\begin{eqnarray}
I_{Fe}(E) &=&  A_{Fe} (2\pi \sigma_{Fe}^{2})^{-1/2} exp[-(E - E_{Fe})^{2}/2\sigma_{Fe}^{2}] \nonumber
\end{eqnarray}


%
%

\vspace{10 mm}

The model (referred to as the PC-Model) consists of an unabsorbed continuum component 
plus an absorbed component plus an iron emission line. 
Both continuum components have the same spectral parameters except for normalization. 
We first fitted all the spectra with all parameters free except the iron line width (FWHM) 
was fixed at 0.6 keV, and the smoothing function as mentioned above. 
Tests showed that allowing a free iron line width did not improve the fits.
These first fits showed that the spectral parameters are poorly constrained because of 
the large number of free spectral parameters. 
So we fixed the cutoff and folding energies because they turned out to be consistent, 
within the uncertainties, with a constant for both LS and SH-state spectra. 
The cutoff energy and the folding energy was fixed at the obtained values in the current 
work and are as following : 19.0 and 7.0 for LS; 19.14 and 7.3 for SH-state. the current values are 
lower than the previous results for Her-X1 (Coburn et al. 2002), where they found that the cutoff energy 
and the folding energy are 22.0 +/- 1.4/0.8  and 10.8 +/- 0.2/0.3 respectively. 
Coburn et al. (2002) used out of eclipse data during the peak of Main-High state, which means that
the observed spectrum represented the intrinsic neutron star polar cap emission, however in our case 
we used LS and SH data. The contribution from the disk is high during SH, and scattered emission 
from the corona is important during LS, so in our data we do not measure the intrisic neutron star 
polar cap emission. Our lower value for the folding enrgy than previously obtained for Main-High state 
is likely due to the influence of the cyclotron absorption line at energies below the centroid.
This left us with a total of six spectral parameters: the partial covering column density 
$N_{H}$ and covering fraction $F$ , photon-index, continuum normalization, 
iron-line energy, and line normalization. 

\vspace{10 mm}

Representative spectra from LS (MJD/$\phi_{orb}$: 52261.49/0.2 top-left-panel and 52262.01/0.5 
top-right-panel) and SH-state (MJD/$\phi_{orb}$: 52264.84/0.17 bottom-left-panel and 52265.25/0.41 
bottom-right-panel) are shown in figure (3). 
Generally the presence of a prominent iron emission line in LS is clearly seen. 
The line normalization increases from 0.0005 to 0.0011 $photons$ $cm^{-2}$ $s^{-1}$ for  
the two spectra shown in Fig. 3 for LS and increases from 0.0035 to 0.0044 $photons$ $cm^{-2}$ $s^{-1}$ for  the 
two spectra shown for SH-state. 

\vspace{10 mm}

The time dependence of the spectral parameters is shown in Figure (4) for both LS (left panels) 
and SH-state (right panels). 
In LS, $N_{H}$ and iron $K_{\alpha}$ normalization are most strongly modulated 
with orbital phase and both are maximum around orbital phase 0.5. 
The iron line energy decreases and the power-law index decreases (harder spectrum)
around orbital phase 0.5. 
The increase in $N_{H}$, increase in iron line normalization and decrease in iron line
energy and power-law index are all consistent with reflected emission from the companion.
For SH state, modulations for most parameters is seen. 
Similar to LS, the spectra are harder (smaller power-law index) around $\phi_{orb} = 0.5$. 
The energy of the iron $K_{\alpha}$ line varies with a minimum value at $\phi_{orb} \sim 0.5$ , 
while the iron line normalization increases at $\phi_{orb} \sim 0.5$. 
The modulations of iron line normalization, iron line energy  and 
power-law index are all consistent with having a component which is reflection of X-rays off the companion.

The power-law (continuum) normalization shows a weak variation in LS 
and a dip around $\phi_{orb} \sim 0.5$ in SH-state (Fig. 4), 
but is not an accurate measure of the continuum strength because 
the power-law index is also variable and the flux in the continuum depends on both parameters. 
Rather, to measure the change in continuum flux we calculated the flux using the $flux$ command in XSPEC.
This calculates the absorption-corrected fluxes of the model used within a given energy range.
We consider the flux in several energy ranges: 3-5 keV, 8-10 keV, 10-15 keV, 15-20 keV, 20-30 keV, and 10-30 keV.
This gives a good overview of how the continuum is behaving 
(see Fig. 5 "left panels" for LS fluxes, and "right panels" for SH-state). 
The continuum flux has a weak time dependence in the 3-5 keV band 
but a strong peak around orbital phase 0.5 for the bands above 10 keV, 
and intermediate behaviour for the 8-10 keV band. 
In LS, the behaviour of the continuum flux is consistent with a significant contribution
of reflected X-rays around orbital phase 0.5, 
where the reflected X-rays off the face of HZ Her 
are expected to have a harder spectrum than the incident X-rays (e.g. Fig. 2 of Basko et al. 1974).
This is consistent with Figure 5: the modulation vs. orbital phase increases
as energy is increased, similar to what is expected.
The 3-5 keV band is least modulated and the 20-30 keV band is most modulated.
However in addition to the predicted modulation from X-ray reflection from the companion
star, all bands include a roughly constant contribution. 
In particular, Fig. 1b of Basko et al. 1974 shows the reflection contribution lower by a 
factor of 10 at orbital phases 0.2 and 0.8, compared to the peak at orbital phase 0.5.
But the data of Fig. (5) here for LS show the 3-5 keV data lower by a factor of only $\sim2$
and the 20-30 keV data lower by a factor of $\sim4$.
Thus there is another component in addition to the reflection component for LS. 
We associate this component with the X-ray corona of Her X-1 (Leahy \& Yoshida 1994; Kuster et al. 2005; and Leahy 2015).
For SH the modulation is weaker than for LS. 
This is caused by having an additional strong component, the emission from the neutron star 
and from the inner accretion disk (Leahy, 2002).

\vspace{10 mm}

To further investigate the contribution of reflection plus corona X-rays to the SH state, 
we compare the shapes of the fluxes between LS and SH in the 6 energy ranges  
(3-5, 8-10, 10-15, 15-20, 20-30 and 10-30 keV). 
Fig. 6 shows a fit to the LS fluxes (left panels). In the middle and right panels,
the LS flux plus a constant is fit to the second and third orbital periods fluxes 
of the SH-state (in the same energy band).
The $\chi^2$ values for the fits vary from 300 to 800 for 25 to 27 degrees of freedom, 
except for the 3-5 keV band which has large error bars and has $\chi^2$ of 50 and 59 
for 25 and 27 degrees of freedom.
It is clear that the SH fluxes cannot be represented by a constant plus the LS fluxes. 

\vspace{10 mm}

Reflected X-rays off of HZ Her are not significant below 5 keV (e.g. Basko et al. 1974). 
Thus it is reasonable to use the 3-5 keV flux as a measure of the 35-day time dependence
of the X-rays scattered from the inner disk.
I.e. we scale the 3-5 keV flux by a constant to represent the inner disk scattered X-ray flux 
as a function of time for higher energy ranges.
The flux above 10 keV has a small but significant contribution from reflected X-rays off HZ Her. 
The neutron star emission is known to not change with 35-day phase, but rather the disk is precessing slowly. 
This implies the X-ray illumination of HZ Her is almost the same for late LS (35-day phase 0.51-0.56, Fig. 1) 
and early SH (35-day phase 0.57-0.74),  and the reflected
X-ray emission should be nearly the same for LS and SH.
We use the $>$10 keV fluxes in LS to represent the contribution of reflected 
X-rays to the $>$10 keV fluxes in SH.
Thus the two component models for the flux in band $X$ are:  
\begin{equation}
f1_X(t) = C1 \times f_{3-5keV}(t) + C2 \times f_{X,LS}(\phi_{orb})
\end{equation}
and
\begin{equation}
f2_X(t) = C1 \times f_{3-5keV}(t) + f_{X,LS}(\phi_{orb})
\end{equation}
where $f_{3-5keV}(t)$ is the 3-5 keV flux for the same 
time and $f_{X,LS}(\phi_{orb})$ is the LS flux for the same energy band and
same orbital phase.
For each of the four orbits in SH, these two models were fit to the SH data in the 
bands with energy $>$10 keV. 
The fit results ($\chi^2$) show that the extra parameter $C2$ in the first model is not needed.
The $\chi^2$ for both models and the best fit parameter $C1$ for the second model are given in Table (1). 
The best fit of the second model is shown in Fig. 7 for the 10-30 keV band, which has the best statistics.
We find that the 10-30 keV flux is well represented by this two component model for all four SH orbits. 
This demonstrates that this two component model represents the data well.

\section{Discussion}

The spectral parameters in Fig. 4 and absorption-corrected fluxes in Fig. 5 
show that the LS emission of Her X-1 is composed of reflected X-rays off of HZ Her 
and a component which we attribute to scattered X-rays from the corona.
The  reflected X-rays are orbital-phase dependent, and the scattered X-rays are nearly constant. 
The expected intensity of the scattered X-rays in the 9-20 keV band is $\sim$0.003 
c/s/cm$^{2}$ (Fig. 4 in Leahy 2015). 
This is consistent with the values here (Fig. 5) for the constant part of the emission 
during LS in the 10-15 keV  band ($\sim$0.0017 c/s/cm$^{2}$) plus 
15-20 keV  band ($\sim$0.0013 c/s/cm$^{2}$).

The column density in LS (Fig. 4) shows modulation with orbital phase with $N_{H}$ varying 
between $30 \times 10^{22}$ and $100 \times 10^{22}$ $ cm^{-2}$,
 with highest values attained around orbital phases $ \phi_{orb} = 0.58$. 
 The covering fraction, however, is nearly constant. 
Both results conform with results obtained earlier by Mihara et al. 1991 \& Coburn et al. 2000. 
The absorbed component in LS at early and late orbital phases, when reflection off of HZ Her 
should be small, is likely the neutron star emission absorbed by the inner disk, 
then scattered by the extended corona, which was clearly detected by Leahy (2015). 
The unabsorbed component, is probably emission from the neutron star which is scattered by
the corona. 

\vspace{10 mm}

During SH, eclipse of the X-ray source by HZ Her (Leahy, 2000) shows that the source is
extended in size, consistent with the inner accretion disk. 
In models for the 35-day cycle light curve (Leahy 2002) and for the 35-day evolution
of the pulse profile (Scott, Leahy and Wilson 2000) the inner accretion disk is expected 
to have a significant contribution to the total flux emitted from the system.
However the time dependence of the disk emission is not well-determined yet, partly
because it is sensitive to the disk geometry which determines the illumination by the 
neutron star and the disk self-shadowing. Thus, we assumed that the disk emission is 
best decribed by the 3-5 keV flux to scale the emission in higher energy bands 
($>$ 10 keV) and used the reflected plus corona emission during LS to model the flux 
during SH-state. We found that the scaling factor for the 4 orbits in SH fell in 
a narrow range of 1.41 to 1.53 with $\chi^2$ values ranging from 4.8 to 22 for 9 to 27 $dof$. 
This strongly supports the statement that the X-ray reflection flux of LS also appears during SH-state.
 
\vspace{10 mm}

This is the first time in which the signature of X-rays reflected off of the inner face of 
HZ Her is found during the SH-state. 
We find the evidence in orbital modulations of the spectral parameters (Fig. 4) 
and of the flux (Fig. 5).  
Leahy (2000) studied Ginga observations in the middle part of SH-state 
(MJD 47664.3 to 47664.6 or orbital phase $\sim$ 0.31 to 0.49) which shows slowly increasing 
count rates in all energy bands. This might be caused by atmospheric reflection, 
but orbital phase coverage of that data was too small to draw that conclusion.

\vspace{10 mm}

From both Fig. 4 and Fig. 5, during LS there is a clear asymmetry in 
spectral parameters (Fe line norm and $N_H$) and flux about orbital phase 0.5. 
The maximum reflected intensity is seen at orbital phase $\sim0.55$. 
Similar asymmetry has been observed in the optical emission lines (Quaintrell et al.2001) 
and continuum/emission-lines in UV band (Vrtilek \& Cheng 1996 ; Vrtilek et al. 2001).
This asymmetry is an indication that the irradiation on the inner face of HZ Her is not uniform 
where the leading side is brighter in X-rays than that of the trailing side. 
The non-uniformity of the irradiation on the inner face of HZ Her could be caused by 
shadowing by either accretion disk and/or stream.
An asymmetry is expected from the disk shadow of the X-ray illumination of HZ Her 
(Fig. 5 of Leahy, Marshall and Scott 2000).
We recall that the disk precession is retrograde to the binary orbit, so that the disk
shadow move across the face of HZ Her with the beat period of 1.65 days. 
This means that the shadowing can change significantly between orbital phase 0.4 and
0.6, and cause the observed asymmetry in reflected emission that change with 35-day phase.
On the other hand, shadowing by accretion stream is not expected to change with 35-day phase. 
More observations covering different orbital periods during LS throughout 35-day cycle is needed before 
making a conclusive statement about the cause of this asymmetry. 
For SH the asymmetry is not easy to see because of the variability in the disk plus 
neutron star emission component.
However, the asymmetry is probably still present, as shown by Fig. 7, which uses the 
asymmetric observed reflection component from LS to fit the SH data.
We note that the changing disk orientation relative to the observer with 35-day phase
gives a changing disk plus neutron star emission.
The changes are most prominent during the first SH orbit (see the 3-5 keV flux in
Fig. 5) caused by the uncovering of the central source from behind the outer edge of the disk. 
The third and fourth SH orbits show the decay of the SH-state for which the inner  
edge of the disk progressively covers of the central source (Scott, Leahy and Wilson 2000).

\section{Conclusion} 

In summary, the spectral parameters in Fig. 4  and fluxes in Fig. 5 
show that the LS emission of Her X-1 is composed of reflected X-rays off of HZ Her 
and scattered X-rays from the corona. 
Due to the presence of a strong direct emission from the neutron-star/accretion-disk 
the reflected component is less prominent during SH-state. 
However modulation of both spectral parameters and high energy fluxes ($>$ 10 keV), 
and the fitting of SH-state hard-fluxes ($>$ 10 keV) with the 
LS fluxes plus scaling the disk by the 3-5 keV flux confirm the presence of 
reflected emission. Asymmetry in both spectral parameters and fluxes is clearly seen 
in LS, and likely present during SH-state but difficult to be distinguished due to 
the strong direct emission. This asymmetry is probably due to the lack of irradiation 
uniformity over the inner face of HZ Her due to shadowing from the accretion-disk and/or stream. 

\acknowledgments
This work is supported by a grant from the Natural Sciences and Engineering Research Council
of Canada to D.L.

\clearpage

\begin{figure}
\epsscale{1.0}
\plotone{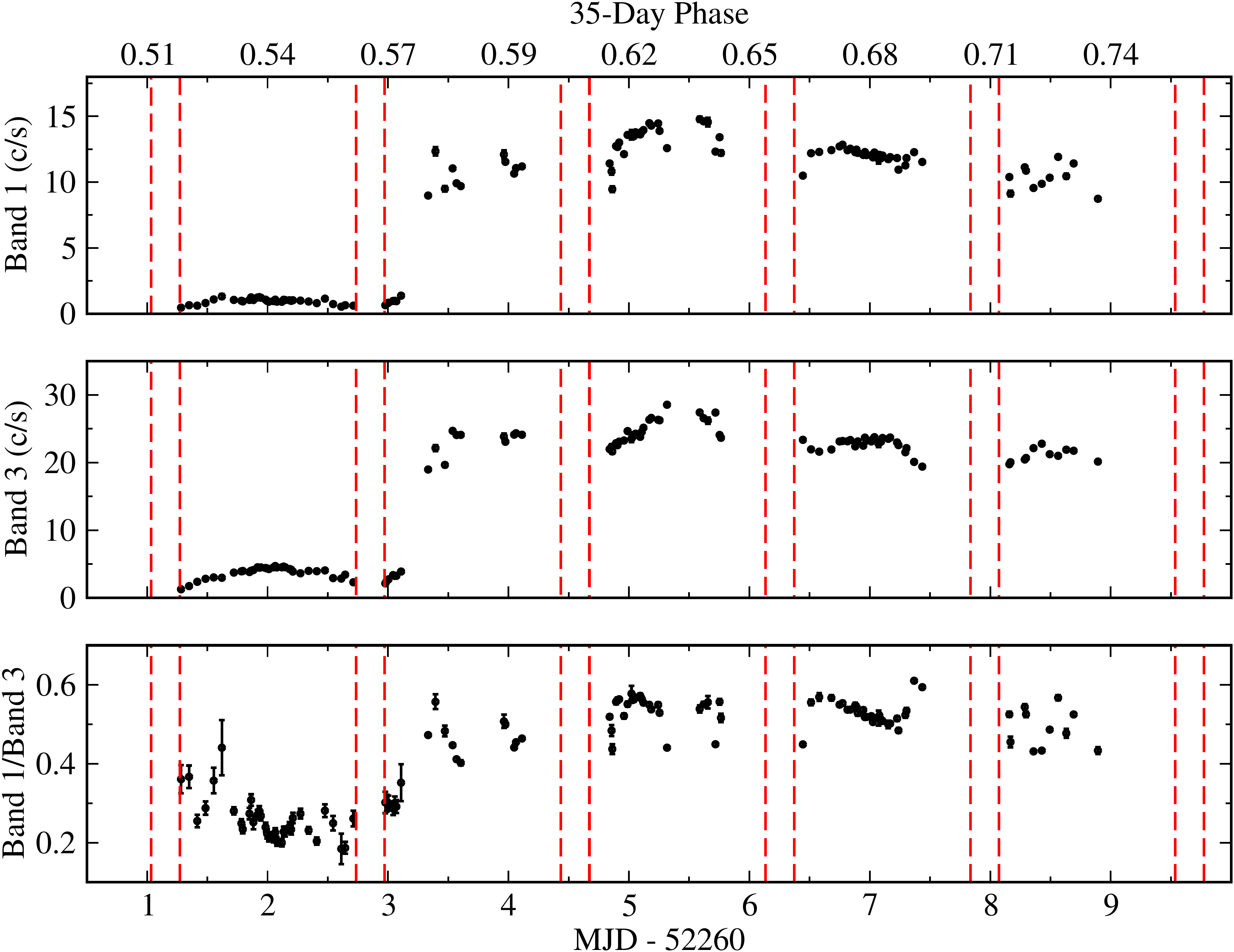}
\caption{RXTE/PCA X-ray light curve: (top) Band 1 2-4 keV, (Middle) Band 3 9-20 keV, (bottom) softness ratio . 
Vertical dashed lines indicate  eclipse ingress and egress. For these light curves we averaged the count rates over time intervals between 112-s and 1680-s with a mean value of $\sim$ 1 ks  \label{fig1}}
\end{figure}

\clearpage

\begin{figure}
\epsscale{1.0}
\plotone{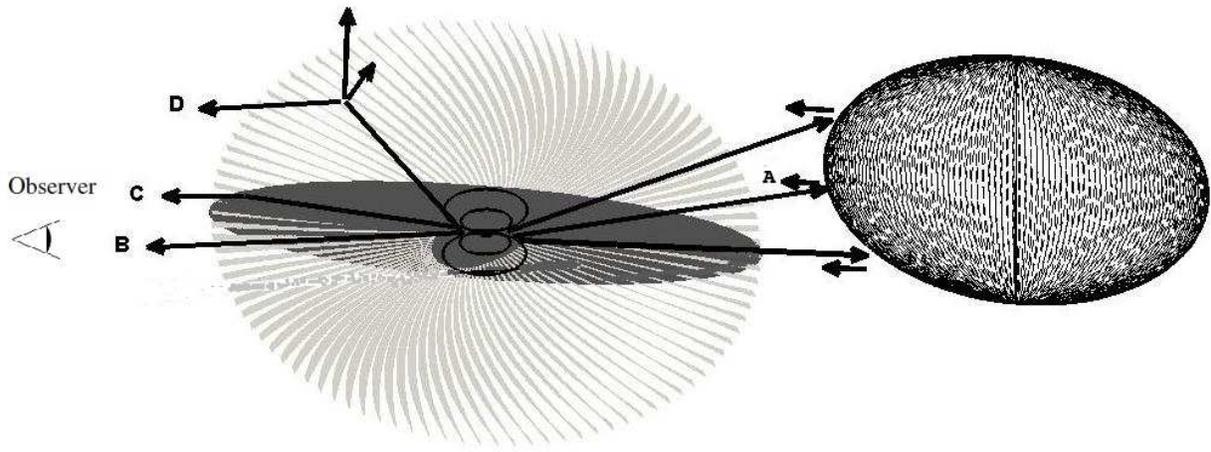}
\caption{Graphical representation of the various emission components in the system Her X-1/HZ Her. 
Emission components are : (A) reflected emission by the companion; (B) Direct emission from the neutron star; 
(C) absorbed/reflected emission by the disk; and (D) scattered emission by the corona. \label{fig2}}
\end{figure}

\clearpage

\begin{figure}
\epsscale{1.0}
\plotone{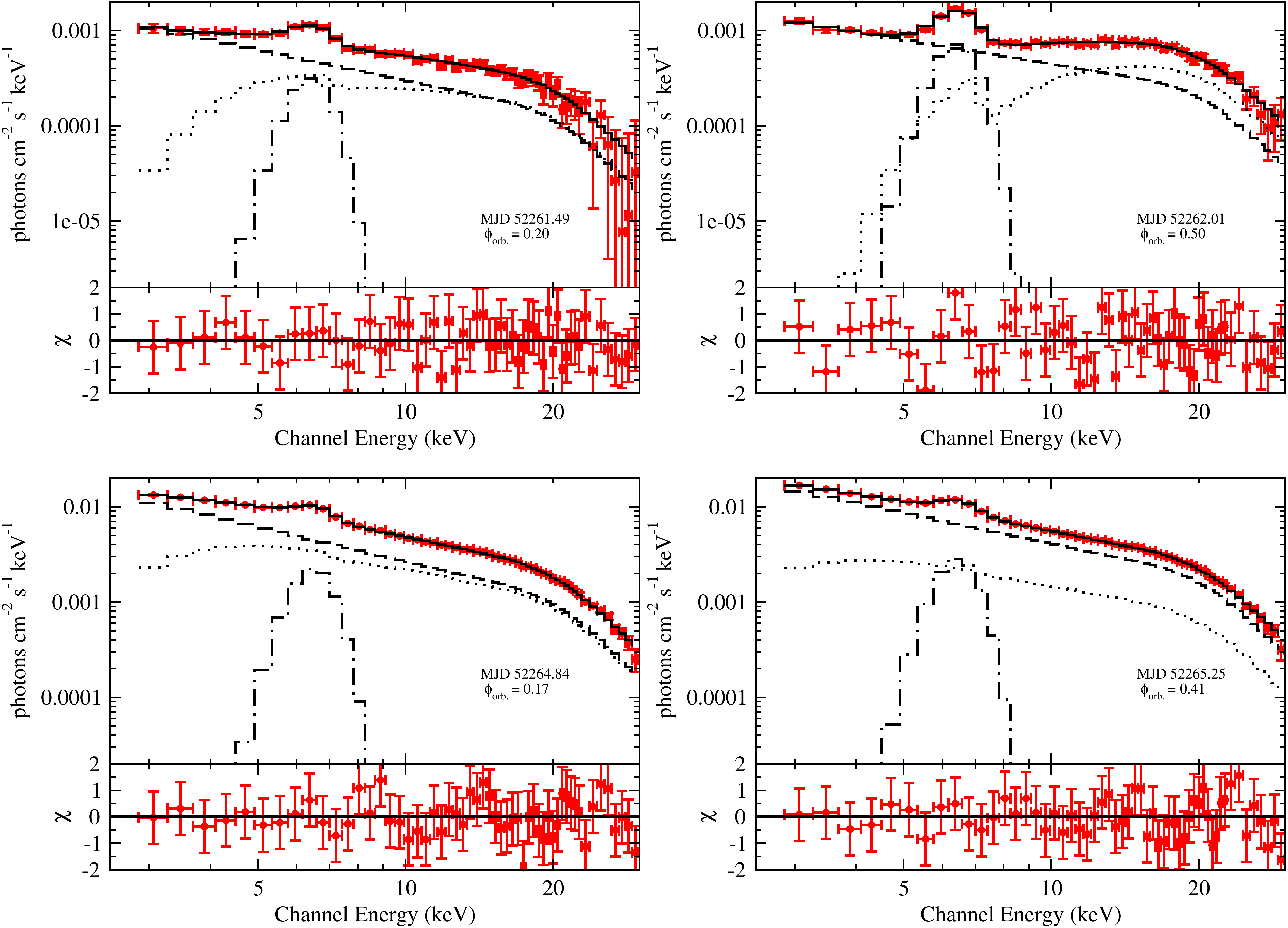}
\caption{Example RXTE/PCA spectra of Her X-1 during low state (top panels), and short-high state (bottom panels).
The observed spectra are points with error bars and the histograms are the model: the absorbed component (dotted line), unabsorbed component (dashed line), iron line (dash-dotted line), and sum of model components (solid line). The lower part of each panel shows the residuals in terms of sigma between the data and the model.
 \label{fig3}}
\end{figure}

\clearpage

\begin{figure}
\epsscale{1.0}
\plotone{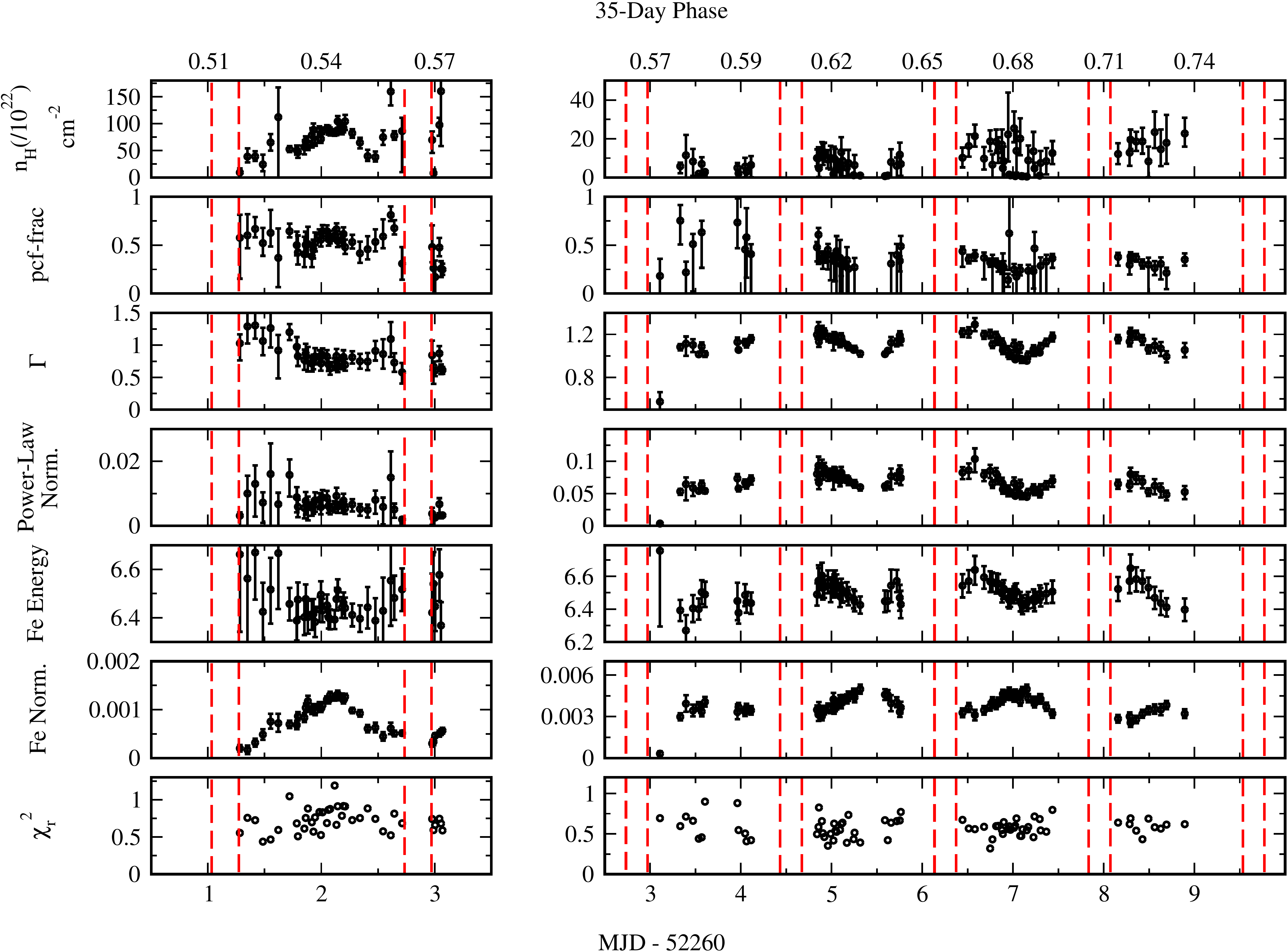}
\caption{Spectral parameters variation of PC-model during low state (left panels), and short high state (right panels). From top to bottom panels: partial covering hydrogen column densities, the covering fraction, photon index $\Gamma$, 
the power-law normalization in units of photons $cm^{-2}$ $s^{-1}$ $keV^{-1}$, Iron line energy in keV, Iron line normalization in units of photons $cm^{-2}$ $s^{-1}$, and reduced chi-squared of the fits.  \label{fig4}}
\end{figure}

\clearpage

\begin{figure}
\epsscale{1.0}
\plotone{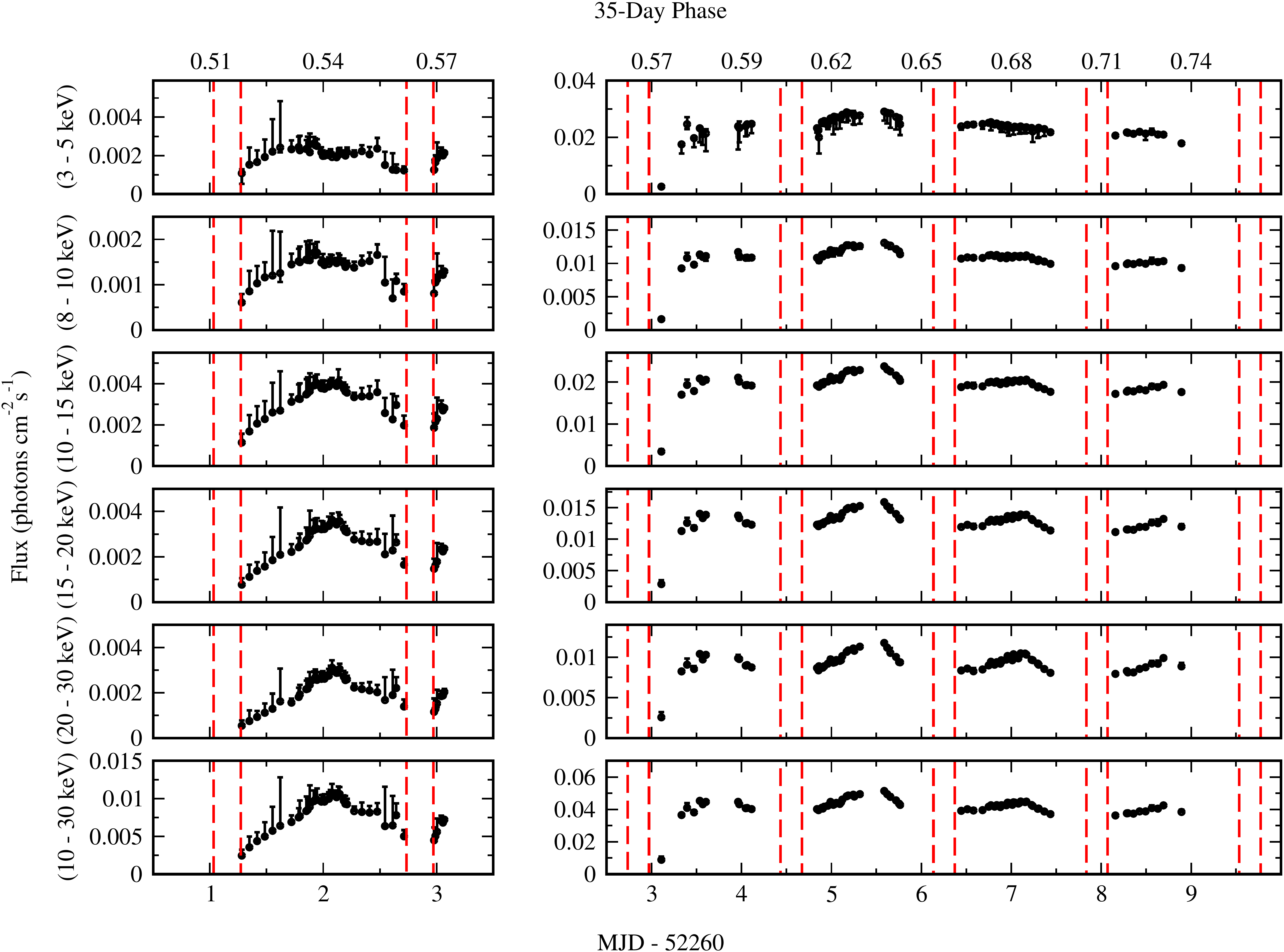}
\caption{The fluxes in different energy ranges during low state (left panels), and short high state (right panels). 
The hard ($h\nu$ $>$ 10 keV) X-ray flux is clearly modulated with orbital phase.  \label{fig5}}
\end{figure}

\clearpage

\begin{figure}
\epsscale{1.0}
\plotone{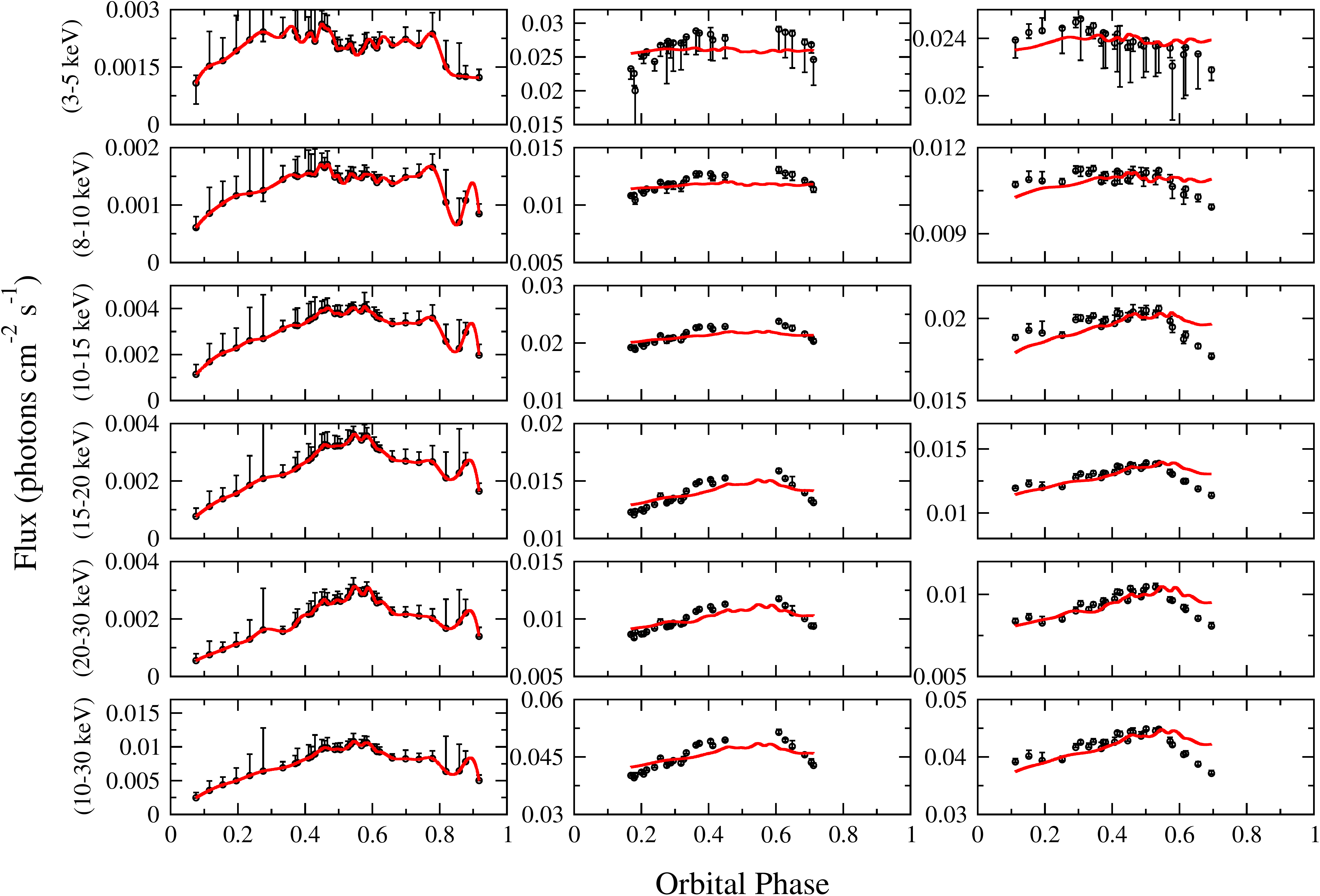}
\caption{Comparison of LS and SH fluxes in the energy bands 3-5, 8-10, 10-15, 15-20, 20-30
and 10-30 keV (top to bottom rows, respectively). 
Left panels: spline fit (solid line) to the LS fluxes (points with error bars); 
middle  panels: fit of the LS fluxes  plus a constant (solid line) to the SH fluxes (points with error bars) for the second SH orbit; 
right panels: fit of the LS fluxes plus a constant (solid line) to the SH fluxes (points with error bars) for the third SH orbit. \label{fig6}}
\end{figure}

\clearpage

\begin{figure}
\epsscale{1.0}
\plotone{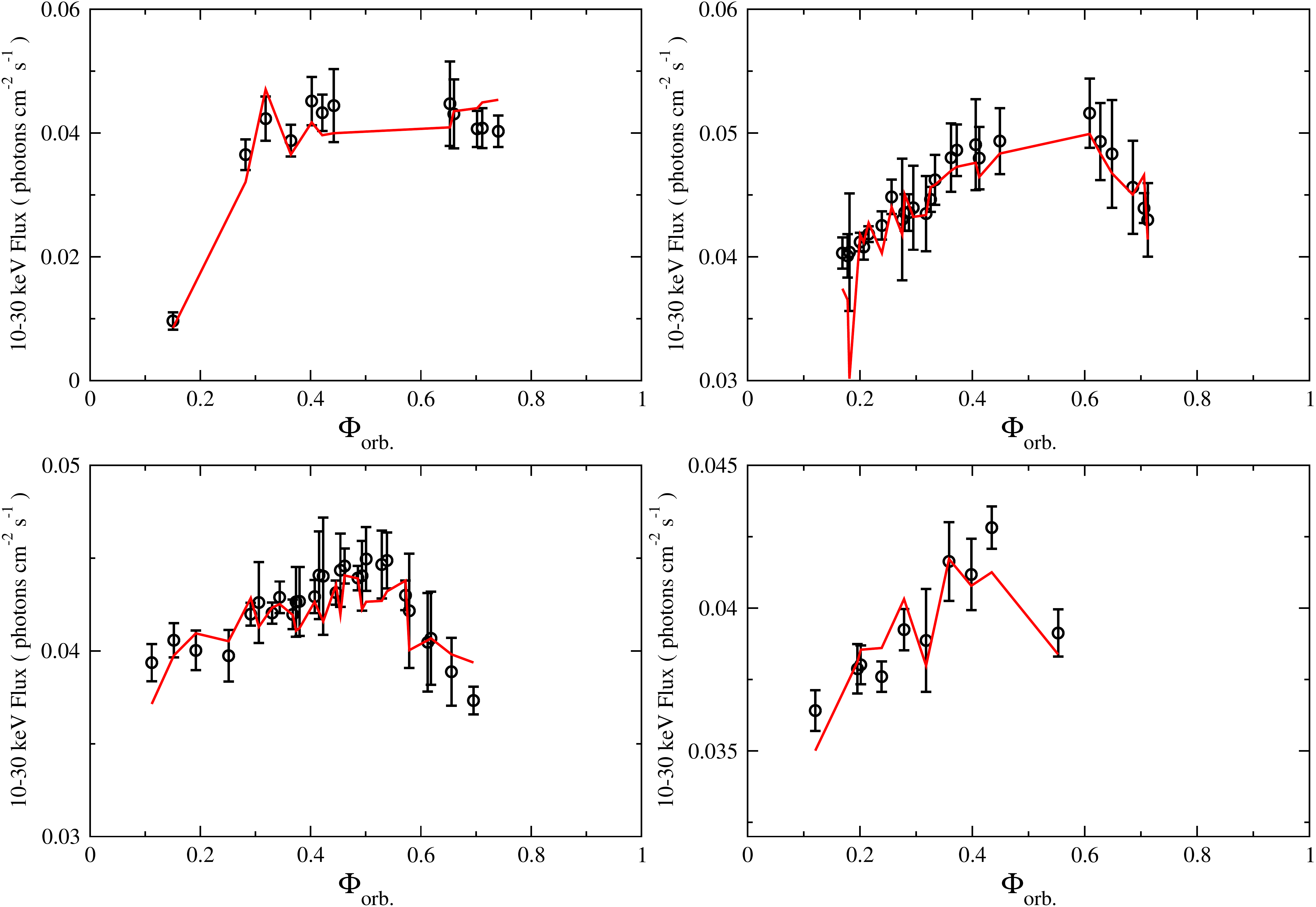}
\caption{Two component model fit to the fluxes in SH for the energy band 10-30 keV, for 
the four orbits of SH data: first orbit- upper left; second orbit- upper right; third orbit-
lower left; fourth orbit- lower right.
In each case, the solid line is the fit, and the error bars include the uncertainties in the 3-5 keV and 10-30 keV data. \label{fig7}}
\end{figure}

\tiny

\clearpage

\renewcommand{\thefootnote}{\alph{footnote}}

\begin{table}[ht]
\begin{center}
\caption{Sample spectral parameters of PCA observations of Her X-1/HZ Her for data shown in figure(3)}
\begin{tabular}{l c c c c}
\hline\hline

Parameter & $Low state^{(a)}$ & $Low state^{(b)}$ & $SH-state^{(c)}$ & $SH-state^{(d)}$
\\
\hline
Hydrogen column density  & $24.2_{-12.7}^{+18.1}$ & $86.9_{-6.4}^{+6.2}$ & $10.0_{-4.4}^{+4.5}$ & $6.3_{-9.6}^{+5.3}$ \\
($10^{22}$  $cm^{-2}$)  & & & & \\

Partial covering fraction & $0.52_{-0.13}^{+0.16}$ & $0.65_{-0.05}^{+0.06}$ & $0.48_{-0.15}^{+0.05}$ & $0.27_{-0.73}^{+0.10}$ \\

Photon index & $1.07_{-0.22}^{+0.21}$ & $0.85_{-0.10}^{+0.10}$ & $1.20_{-0.04}^{+0.03}$ & $1.07_{-0.04}^{+0.03}$ \\

Power-law normalization & $0.72_{-0.63}^{+0.32}$ & $0.88_{-0.31}^{+0.23}$ & $8.03_{-0.88}^{+0.73}$ & $6.61_{-0.72}^{+0.48}$ \\
 ($10^{-2}$ $photons$ $cm^{-2}$ $s^{-1}$ $keV^{-1}$)  & & & & \\

Iron line energy (keV) & $6.42_{-0.13}^{+0.12}$ & $6.43_{-0.05}^{+0.05}$ & $6.49_{-0.07}^{+0.07}$ & $6.45_{-0.06}^{+0.06}$ \\

Iron line normalization & $0.49_{-0.11}^{+0.11}$ & $1.09_{-0.07}^{+0.07}$ & $3.49_{-0.31}^{+0.31}$ & $4.38_{-0.35}^{+0.35}$ \\
 ($10^{-3}$ $photons$ $cm^{-2}$ $s^{-1}$ $keV^{-1}$)  & & & & \\

3-5 keV X-ray flux   & $0.192_{-0.002}^{+0.092}$ & $0.199_{-0.001}^{+0.023}$ & $2.323_{-0.153}^{+0.009}$ & $2.753_{-0.308}^{+0.019}$ \\
($10^{-2}$ $photons$ $cm^{-2}$ $s^{-1}$) & & & & \\

10-30 keV X-ray flux   & $0.497_{-0.012}^{+0.190}$ & $0.959_{-0.002}^{+0.099}$ & $4.017_{-0.013}^{+0.039}$ & $4.796_{-0.034}^{+0.038}$ \\
($10^{-2}$ $photons$ $cm^{-2}$ $s^{-1}$) & & & & \\

$\chi^{2}_{r}$  (49 d.o.f.) & 0.44 & 0.83 & 0.50 & 0.52 \\
\hline
\end{tabular}
  \tablenotetext{(a)}{MJD 52261.49 , $\phi_{orb.}$ = 0.20}
  \tablenotetext{(b)}{MJD 52262.01 , $\phi_{orb.}$ = 0.50}
  \tablenotetext{(c)}{MJD 52264.84 , $\phi_{orb.}$ = 0.17}
  \tablenotetext{(d)}{MJD 52265.25 , $\phi_{orb.}$ = 0.41}

\end{center}
\end{table}

\clearpage

\renewcommand{\thefootnote}{\alph{footnote}}

\begin{table}[ht]
\begin{center}
\caption{Results of model fits to 10-30 keV fluxes in SH-state}
\begin{tabular}{c c c c c}
\hline\hline
SH-Orbit & No. of data & $\chi^{2~(a)}$ & $\chi^{2~(b)}$ & C1$^{(b)}$  
\\
\hline
1 & 12 & 14.2 & 14.6 & 1.61 \\
2 & 26 & 31.7 & 32.8 & 1.46 \\
3 & 28 & 28.9 & 29.8 & 1.45 \\
4 & 10 & 14.0 & 15.3 & 1.54 \\
\hline
\end{tabular}
  \tablenotetext{(a)}{Applies to model of equation 2.}
  \tablenotetext{(b)}{Applies to model of equation 3.}
\end{center}
\end{table}

\end{document}